\documentclass[prl,twocolumn,groupedaddress,showpacs,nofootinbib]{revtex4}
\usepackage{graphicx}
\usepackage{dcolumn}
\usepackage{bm}
\usepackage{amssymb}
\usepackage{mathrsfs}
\usepackage{amsmath}
\usepackage{epsfig}
\usepackage[dvips]{color}
\usepackage{hhline}

\begin{document}

\title{Visible and Dark Matter Genesis and Cosmic Positron/Electron Excesses }

\author{Pei-Hong Gu$^{1}_{}$}
\email{pgu@ictp.it}

\author{Utpal Sarkar$^{2,3}_{}$}
\email{utpal@prl.res.in}

\author{Xinmin Zhang$^{4}_{}$}
\email{xmzhang@ihep.ac.cn}

\affiliation{$^{1}_{}$The Abdus Salam International Centre for
Theoretical Physics, Strada Costiera 11, 34014 Trieste, Italy\\
$^{2}_{}$Physical Research Laboratory, Ahmedabad 380009, India\\
$^{3}_{}$Department of Physics and McDonnell Center for the Space
Sciences, Washington University, St. Louis, MO 63130, USA\\
$^{4}_{}$Institute of High Energy Physics, Chinese Academy of
Sciences, Beijing 100049, China}

\begin{abstract}

Dark and baryonic matter contribute comparable energy density to the
present Universe. The dark matter may also be responsible for the
cosmic positron/electron excesses. We connect these phenomena with
Dirac seesaw for neutrino masses. In our model (i) the dark matter
relic density is a dark matter asymmetry generated simultaneously
with the baryon asymmetry so that we can naturally understand the
coincidence between the dark and baryonic matter; (ii) the dark
matter mostly decays into the leptons so that its decay can
interpret the anomalous cosmic rays with positron/electron excesses.

\end{abstract}


\maketitle

Precise cosmological observations indicate that dark and baryonic
matter have different properties but contribute comparable energy
density to the present Universe. This intriguing coincidence
inspires us to propose a common origin for the creation and
evolution of the dark and baryonic matter. Since the baryonic matter
currently exists because of a matter-antimatter asymmetry, the dark
matter relic density may also be an asymmetry between the dark
matter and dark antimatter \cite{kuzmin1997,kl2004,cht2005,klz2009}.
The dark matter asymmetry and the baryon asymmetry may originate
from decays of the same heavy particles, so that their coincidence
is not surprising at all. On the other hand, recently cosmic-ray
data \cite{chang2008,torii2008,adriani2008,aharonian2008,abdo2009}
suggest \cite{mpsv2009} that (i) the dark matter should dominantly
annihilate or decay into leptons with a large cross section or a
long life time; (ii) the dark matter annihilation or decay should be
consistent with the constraints from the observations on the gamma
and neutrino fluxes. In this paper we explain these phenomena in a
unified scenario where the neutrino masses originate through the
Dirac seesaw mechanism \cite{gh2006}.

\vspace{2mm}

We extend the standard model (SM) with a global $U(1)^{}_{lepton}
\times U(1)^{}_{dark}$ symmetry and include additional particles:
singlet right-handed neutrino
$\nu_{R}^{}(\mathbf{1},\mathbf{1},0)(1,1)$, heavy doublet scalar
$\eta(\mathbf{1},\mathbf{2},-1/2)(0,-1)$, charged singlet scalar
$\xi(\mathbf{1},\mathbf{1},1)(-2,0)$, and neutral singlet scalars
$\sigma(\mathbf{1},\mathbf{1},0)(0,-1)$ and
$\chi(\mathbf{1},\mathbf{1},0)(-2,-1)$, where the transformations
are given under the SM gauge group $SU(3)_{c}^{}\times SU(2)_L^{}
\times U(1)_{Y}^{}$ and the global symmetry $U(1)_{lepton}^{}\times
U(1)_{dark}^{}$. For simplicity, we only present the relevant part
of the Lagrangian for purpose of demonstration,
\begin{eqnarray}
\mathcal{L} &\supset&
-y\overline{\psi_{L}^{}}\eta\nu_{R}^{}-f\xi\overline{\psi_{L}^{c}}
i\tau_2^{}\psi_{L}^{}-\mu\sigma\eta^\dagger_{}\phi-\kappa\chi^\dagger_{}\xi\eta^{T}_{}i\tau_2^{}\phi\nonumber\\
[2mm]
&&+\textrm{H.c}-M_\eta^2\left(\eta^\dagger_{}\eta\right)-m_\xi^2\left(\xi^\dagger_{}\xi\right)-m_\chi^2\left(\chi^\dagger\chi\right)\nonumber\\
[2mm]
&&-\left(\chi^\dagger_{}\chi\right)\left[\alpha\left(\xi^\dagger_{}\xi\right)
+\beta\left(\phi^\dagger_{}\phi\right)+\gamma\left(\sigma^\dagger_{}\sigma\right)\right]\nonumber\\
[2mm]
&&-\left(\sigma^\dagger_{}\sigma\right)\left[\zeta\left(\xi^\dagger_{}\xi\right)
+\epsilon\left(\phi^\dagger_{}\phi\right)+\vartheta\left(\sigma^\dagger_{}\sigma\right)\right]\,.
\end{eqnarray}
Here $\psi_{L}^{}(\mathbf{1},\mathbf{2},-1/2)(1,0)$ and
$\phi(\mathbf{1},\mathbf{2},-1/2)(0,0)$, respectively, are the SM
lepton and Higgs doublets. The right-handed neutrinos neither have
Yukawa couplings with the SM Higgs doublet nor have Majorana masses
as a result of the $U(1)_{lepton}^{}\times U(1)_{dark}^{}$
conservation. The global symmetry $U(1)_{lepton}^{}$ will be exactly
conserved at any energy scales while the global symmetry
$U(1)_{dark}^{}$ will be spontaneously broken above the weak scale.

The singlet scalar $\sigma$ acquires a vacuum expectation value
(VEV) to break $U(1)_{dark}^{}$ and then induces a small VEV of the
heavy doublet scalars $\eta$ after the electroweak symmetry breaking
by $\langle\phi\rangle\simeq 174\,\textrm{GeV}$,
\begin{eqnarray}
\langle\eta\rangle \simeq
-\frac{\mu\langle\sigma\rangle\langle\phi\rangle}{M_{\eta}^2}~~\textrm{for}~~M_{\eta}^{}\gtrsim
\mu \gg \langle\sigma\rangle > \langle\phi\rangle\,.
\end{eqnarray}
Therefore through the Dirac seesaw mechanism \cite{gh2006},
the neutrinos obtain very
small Dirac masses naturally through their Yukawa couplings to the heavy
doublet,
\begin{eqnarray}
m_{\nu}^{}= y\langle\eta\rangle\,.
\end{eqnarray}

\vspace{2mm}

\begin{figure*}
\vspace{8.0cm} \epsfig{file=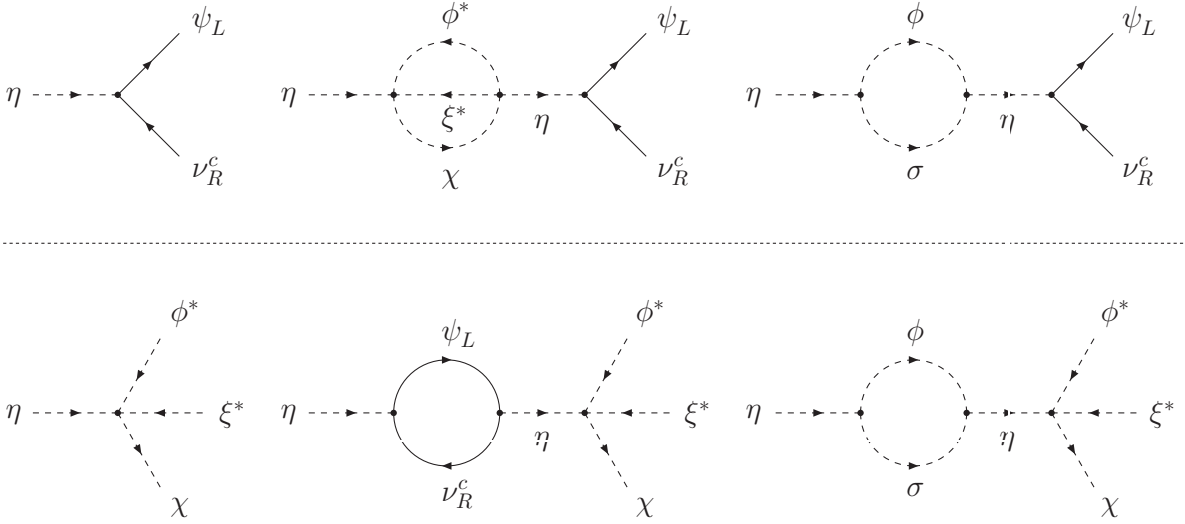, bbllx=6.3cm, bblly=5.9cm,
bburx=16.3cm, bbury=15.9cm, width=8cm, height=8cm, angle=0, clip=0}
\vspace{-9.0cm} \caption{\label{asymmetry} The lepton number
conserving decays for generating the desired lepton asymmetry and
dark matter asymmetry.}
\end{figure*}

In our model, the heavy doublet scalar $\eta$ has three lepton
number conserving decay channels:
\begin{eqnarray}
\eta\,\rightarrow\, \psi_{L}^{}\nu_{R}^{c} \,,\quad
\eta\,\rightarrow\, \phi^\ast_{}\xi^\ast_{}\chi\,,\quad
\eta\,\rightarrow\, \phi\,\sigma \,.
\end{eqnarray}
We consider two heavy scalars $\eta_{1,2}^{}$ to incorporate CP
violation. These decays of $\eta_{1,2}^{}$ at loop orders (as shown
in Fig. \ref{asymmetry}) can interfere to generate three types of
lepton asymmetries after $\eta_{1,2}^{}$ go out of equilibrium: the
first one (including that transferred from the charged singlet
scalar $\xi$) is stored in the left-handed lepton doublets
$\psi_{L}^{}$; the second one is stored in the singlet right-handed
neutrinos $\nu_{R}^{}$; and the third one is stored in the neutral
singlet $\chi$, although the total lepton asymmetry vanishes as the
lepton number is exactly conserved.

The effective Yukawa couplings of the left-handed lepton
doublets to the SM Higgs doublet and the right-handed neutrinos are
extremely weak, so that they can not go into equilibrium until the
temperature falls well below the electroweak scale. This will
prevent the $\nu_{R}^{}$ asymmetry to cancel the $\psi_{L}^{}$
asymmetry before the sphaleron transitions are over. Thus,
before the electroweak phase transition, the
sphaleron \cite{krs1985} action will partially convert the lepton
asymmetry stored in the left-handed lepton doublets to the baryon
asymmetry in the Universe. Therefore, we can make use of the
leptogenesis \cite{fy1986} via neutrinogenesis \cite{dlrw1999}
mechanism to understand the baryon asymmetry in the Universe. On the
other hand, the asymmetry between the neutral singlet $\chi$ and its
CP-conjugate will always survive after it is produced because there
are no other processes breaking the lepton number stored in $\chi$.
We will show later that this $\chi$ asymmetry corresponds to an
amount of energy density, equal to the dark matter relic
density, so that $\chi$ becomes the dark matter candidate.

\vspace{2mm}

For an estimate of the amount of CP asymmetry, we work in the basis,
in which the $\eta_{1,2}^{}$ mass matrix is real and diagonal:
$M_{\eta}^{2}=\textrm{diag}\left(M_{\eta_1^{}}^{2},
M_{\eta_2^{}}^{2}\right)$. We also assume $M_{\eta_2^{}}^{} \gg
M_{\eta_1^{}}^{}$, so that the final lepton asymmetry stored in the
left-handed lepton doublets $\psi_{L}^{}$ and the dark matter
asymmetry stored in the neutral singlet $\chi$ should mainly come
from the decays of the lighter $\eta_{1}^{}$:
\begin{eqnarray}
\hskip -.4cm\varepsilon_{\eta_1^{}}^{\psi_{L}^{}}&\equiv&
\frac{\Gamma(\eta_{1}^{} \rightarrow\psi_L\nu^c_R
)-\Gamma(\eta_{1}^{\ast}\rightarrow\psi^c_L\nu_R)}{\Gamma_{\eta_1^{}}^{}}\nonumber\\
[1mm] &&+2\frac{\Gamma(\eta_{1}^{} \rightarrow
\phi^\ast_{}\xi^\ast_{}\chi)-\Gamma(\eta_{1}^{\ast}\rightarrow\phi\xi\chi^\ast_{})}{\Gamma_{\eta_1^{}}^{}}\nonumber\\
[2mm]
&=&\frac{1}{4\pi}\frac{\textrm{Im}\left[\textrm{Tr}\left(y_{1}^{\dagger}y_{2}^{}\right)\left(\frac{\mu_1^\ast\mu_2^{}}{M_{\eta_2^{}}^2}
-\frac{1}{32\pi^2_{}}\kappa_1^{}\kappa_2^\ast
\frac{M_{\eta_1^{}}^2}{M_{\eta_2^{}}^2}\right)\right]}
{\textrm{Tr}\left(y_{1}^{\dagger}y_{1}^{}\right)+\frac{1}{32\pi^2_{}}|\kappa_1^{}|^2_{}+\frac{|\mu_1^{}|^2_{}}{M_{\eta_1^{}}^2}}\nonumber\\
[1mm] &&
-\frac{1}{64\pi^3}\frac{\textrm{Im}\left\{\kappa_1^{\ast}\kappa_2^{}\left[\frac{\mu_1^\ast\mu_2^{}}{M_{\eta_2^{}}^2}
+\textrm{Tr}\left(y_{2}^{\dagger}y_{1}^{}\right)\frac{M_{\eta_1^{}}^2}{M_{\eta_2^{}}^2}\right]\right\}}
{\textrm{Tr}\left(y_{1}^{\dagger}y_{1}^{}\right)+\frac{1}{32\pi^2_{}}|\kappa_1^{}|^2_{}+\frac{|\mu_1^{}|^2_{}}{M_{\eta_1^{}}^2}}\,,
\\
[4mm]\varepsilon_{\eta_1^{}}^{\chi}&\equiv& \frac{\Gamma(\eta_{1}^{}
\rightarrow
\phi^\ast_{}\xi^\ast_{}\chi)-\Gamma(\eta_{1}^{\ast}\rightarrow\phi\xi\chi^\ast_{})}{\Gamma_{\eta_{1}^{}}^{}}\nonumber\\
[1mm]
&=&-\frac{1}{128\pi^3}\frac{\textrm{Im}\left\{\kappa_1^{\ast}\kappa_2^{}\left[\frac{\mu_1^\ast\mu_2^{}}{M_{\eta_2^{}}^2}
+\textrm{Tr}\left(y_{2}^{\dagger}y_{1}^{}\right)\frac{M_{\eta_1^{}}^2}{M_{\eta_2^{}}^2}\right]\right\}}
{\textrm{Tr}\left(y_{1}^{\dagger}y_{1}^{}\right)+\frac{1}{32\pi^2_{}}|\kappa_1^{}|^2_{}+\frac{|\mu_1^{}|^2_{}}{M_{\eta_1^{}}^2}}\,.
\phantom{xxx}
\end{eqnarray}
As the decays of $\xi$ into two leptons $\psi_L$ are in equilibrium,
we included the lepton asymmetry in $\xi$ for estimating the total
lepton asymmetry involved in the sphaleron process. Here
$\Gamma_{\eta_i^{}}^{}$ denotes the total decay width of
$\eta_{i}^{}$ or $\eta_{i}^{\ast}$,
\begin{eqnarray}
\Gamma_{\eta_i^{}}&\equiv&\Gamma\left(\eta_{i}^{}\rightarrow
\psi_{L}^{}\nu_{R}^{c}\right)+\Gamma\left(\eta_{i}^{}\rightarrow
\phi^\ast_{}\xi^\ast_{}\chi^{}_{}\right) +\Gamma\left(\eta_{i}^{}\,
\rightarrow
\phi^{}_{}\sigma^{}_{}\right)\nonumber\\[2mm] &=&
\frac{1}{16\pi}\left[\textrm{Tr}\left(y_{i}^{\dagger}y_i^{}\right)
+\frac{1}{32\pi^{2}_{}}|\kappa_{i}^{}|^{2}_{}+\frac{|\mu_{i}^{}|^{2}_{}}{M_{\eta_{i}^{}}^{2}}\right]M_{\eta_{i}^{}}^{}\,.
\end{eqnarray}
The unitarity and the CPT conservation imply the total decay width
of $\eta_1$ and $\eta^\ast$ to be the same $\Gamma_{\eta_i}
= \Gamma_{\eta_i^\ast}$. For simplicity, we
assume
$y_{2}^{}=y_{1}^{}e^{i\varphi}_{},\,\kappa_{2}^{}=\kappa_{1}^{}e^{i\varphi}_{},\,\mu_{2}^{}=\mu_{1}^{}e^{i(\delta-\varphi)}_{}$
and then derive
\begin{eqnarray}
\varepsilon_{\eta_1^{}}^{\psi_{L}^{}}&=&\frac{\sin\delta}{4\pi}\frac{\left[\textrm{Tr}\left(y_{1}^{\dagger}y_{1}^{}\right)
-\frac{1}{16\pi^2_{}}|\kappa_1^{}|^2_{}\right]\frac{|\mu_1^{}|^2_{}}{M_{\eta_2^{}}^2}}
{\textrm{Tr}\left(y_{1}^{\dagger}y_{1}^{}\right)+\frac{1}{32\pi^2_{}}|\kappa_1^{}|^2_{}+\frac{|\mu_1^{}|^2_{}}{M_{\eta_1^{}}^2}}\,,\\
[1mm]
\varepsilon_{\eta_1^{}}^{\chi}&=&-\frac{\sin\delta}{4\pi}\frac{\frac{1}{32\pi^2_{}}|\kappa_1^{}|^2_{}\frac{|\mu_1^{}|^2_{}}{M_{\eta_2^{}}^2}}
{\textrm{Tr}\left(y_{1}^{\dagger}y_{1}^{}\right)+\frac{1}{32\pi^2_{}}|\kappa_1^{}|^2_{}+\frac{|\mu_1^{}|^2_{}}{M_{\eta_1^{}}^2}}\,.
\end{eqnarray}
The ratio between $\varepsilon_{\eta_1^{}}^{\psi_L^{}}$ and
$\varepsilon_{\eta_1^{}}^{\chi}$ then becomes,
\begin{eqnarray}
\varepsilon_{\eta_1^{}}^{\psi_{L}^{}}:\varepsilon_{\eta_1^{}}^{\chi}=\textrm{Tr}\left(y_{1}^{\dagger}y_{1}^{}\right)
-\frac{1}{16\pi^2_{}}|\kappa_1^{}|^2_{}:-\frac{1}{32\pi^2_{}}|\kappa_1^{}|^2_{}\,.
\end{eqnarray}

\vspace{2mm}

The final baryon asymmetry and dark matter asymmetry would
contribute energy density to the present Universe as below
\cite{kt1990},
\begin{eqnarray}
\label{baryon-asymmetry}
\rho_{B}^{0}&=&n_{B}^{0}m_{N}^{}=\frac{n_{B}^{0}}{s_0^{}}m_{N}^{}s_0^{}
=-\frac{28}{79}\frac{n_{L_{SM}^{}}^{}}{s}\left|_{T\simeq M_{\eta_1^{}}^{}}^{}m_{N}^{}s_0^{}\right.\nonumber\\
&\simeq&
-\frac{28}{79}\varepsilon_{\eta_1^{}}^{\psi_{L}^{}}\frac{n_{\eta_1^{}}^{eq}}{s}\left|_{T\simeq
M_{\eta_1^{}}^{}}^{}m_{N}^{}s_0^{}\right.\,,\\
[1mm] \label{dark-matter-asymmetry}
\rho_{\chi}^{0}&=&n_{\chi}^{0}m_{\chi}^{}=\frac{n_{\chi}^{0}}{s_0^{}}m_{\chi}^{}s_0^{}
=\frac{n_{\chi}^{}}{s}\left|_{T\simeq M_{\eta_1^{}}^{}}^{}m_{\chi}^{}s_0^{}\right.\nonumber\\
&\simeq&
\varepsilon_{\eta_1^{}}^{\chi}\frac{n_{\eta_1^{}}^{eq}}{s}\left|_{T\simeq
M_{\eta_1^{}}^{}}^{}m_{\chi}^{}s_0^{}\right.\,,
\end{eqnarray}
where $m_{N}^{}\simeq 1\,\textrm{GeV}$ is the masses of the
nucleons, $s$ is the entropy density, $n_{B}^{}$ and $n_{\chi}^{}$,
respectively, are the number density of baryon and dark matter,
$n_{\eta_{1}^{}}^{eq}$ is the equilibrium distribution of the heavy
singlet $\eta_{1}^{}$. The solutions
(\ref{baryon-asymmetry}) and (\ref{dark-matter-asymmetry}) for
the baryon and dark matter density are obtained, assuming that
the decays of $\eta_1$ satisfies the out-of-equilibrium condition,
\begin{eqnarray}
\Gamma_{\eta_1^{}}^{}\lesssim H(T)\left.\right|_{T=
M_{\eta_1^{}}^{}}^{}=
\left(\frac{8\pi^{3}_{}g_{\ast}^{}}{90}\right)^{\frac{1}{2}}_{}\frac{M^{2}_{\eta_1^{}}}{M_{\textrm{Pl}}^{}}
\,,
\end{eqnarray}
where $H(T)$ is the Hubble constant with relativistic degrees of freedom
$g_{\ast}^{}\simeq 100$ and the Planck mass
$M_{\textrm{Pl}}^{}\simeq 10^{19}_{}\,\textrm{GeV}$. In the presence
of the fast annihilation between the dark matter and dark antimatter, the
dark matter asymmetry should be equivalent to the dark matter relic
density. In this scenario, the contributions from the baryonic and
dark matter to the present Universe should have the following ratio,
\begin{eqnarray}
\Omega_{B}^{}:\Omega_{\chi}^{}\equiv\rho_{B}^{0}:\rho_{\chi}^{0}=-\frac{28}{79}\varepsilon_{\eta_1^{}}^{\psi_{L}^{}}
m_{N}^{}: \varepsilon_{\eta_1^{}}^{\chi} m_{\chi}^{}\,.
\end{eqnarray}
Conventionally, we define
\begin{eqnarray}
\eta_{B}^{0}&=&\frac{n_B^{0}}{n_{\gamma}^{0}} \simeq
7.04\times\left[-\frac{28}{79}\varepsilon_{\eta_1^{}}^{\psi_{L}^{}}\frac{n_{\eta_1^{}}^{eq}}{s}\left|_{T\simeq
M_{\eta_1^{}}^{}}^{}\right.\right]\nonumber\\
[1mm]
&\simeq&\frac{7.04\,\varepsilon_{\eta_1^{}}^{\psi_{L}^{}}}{15\,g_\ast^{}}\,,
\end{eqnarray}
to describe the current baryon asymmetry. Here $n_{\gamma}^{}$ is
the number density of photon.

\vspace{2mm}

For giving a numerical example, we take $M_{\eta_1^{}}
=0.1M_{\eta_2^{}}^{}=5\times 10^{13}_{}\,\textrm{GeV},\, |\mu_1^{}|
=|\mu_2^{}|=10^{12}_{}\,\textrm{GeV}$ and obtain
$\langle\eta_{1}^{}\rangle =100\langle\eta_{2}^{}\rangle\simeq
1\,\textrm{eV}$ for $\langle\sigma\rangle=1.5\times
10^4_{}\,\textrm{GeV}$. Consequently the neutrino masses would be,
\begin{eqnarray}
\label{parameter-3} m_{\nu}^{}\sim \mathcal{O}(0.01\,\textrm{eV}) ~~
\textrm{for} ~~y_1^{}=y_2^{}e^{-i\varphi}=\mathcal{O}(0.01)\,.
\end{eqnarray}
We further input $m_\chi^{}=4\,\textrm{TeV}$,
$|\kappa_{1}^{}|=|\kappa_{2}^{}|=0.02$, $\sin\delta=-0.5$ and
$\textrm{Tr}\left(m_{\nu}^\dagger
m_\nu^{}\right)=\textrm{Tr}\left(y_{1}^\dagger
y_1^{}\right)\langle\eta_{1}^{}\rangle^2_{}=3\times
10^{-3}_{}\,\textrm{eV}^2_{}$ to obtain a lepton asymmetry
\begin{eqnarray}
\label{parameter-6}
\varepsilon_{\eta_1^{}}^{\psi_L^{}}\simeq-2.3\times 10^3_{}\,
\varepsilon_{\eta_1^{}}^{\chi}\simeq -1.4\times 10^{-7}_{}\,.
\end{eqnarray}
Eventually, we obtain the baryon asymmetry and the dark matter
asymmetry as below
\begin{eqnarray}
\label{parameter-7} \eta_{B}^{0}\simeq 6.2\times
10^{-10}\,,~~\Omega_{\chi}^{}:\Omega_{B}^{}\simeq 5\,,
\end{eqnarray}
which are fully consistent with the experimental observations
\cite{amsler2008}.

In the non-relativistic limit, the annihilation cross-section
of dark matter and dark antimatter reads,
\begin{eqnarray}
\hskip -.5cm\langle\sigma v\rangle =
\frac{1}{32\pi}\left[\left(\alpha-\frac{\gamma\zeta}{2\vartheta}\right)^{2}_{}+2\left(\beta-\frac{\epsilon\zeta}{2\vartheta}\right)^2_{}
+2\gamma^2_{}\right]\frac{1}{m_\chi^2}\,,
\end{eqnarray}
which could be very high as
$\alpha,\beta,\gamma,\zeta,\epsilon,\vartheta<\sqrt{4\pi}$. For
example, we obtain $\langle\sigma
v\rangle=18\,\textrm{pb}\,\left({4\,\textrm{TeV}}/{m_\chi^{}}\right)^2_{}$
for $\alpha,\beta,\vartheta=0.1$ and $\gamma,\zeta,\epsilon=1$. The
thermally produced dark matter, with the mass within the range of a
few GeV to a few TeV, should have an annihilation cross section
slightly smaller than $1\,\textrm{pb}$ to give the desired relic
density (\ref{parameter-7}). If the cross section is much bigger
than $1\,\textrm{pb}$, the thermally produced relic density should
be negligible. Therefore the dark matter relic density can be well
approximated by the dark matter asymmetry.

\vspace{2mm}

\begin{figure}
\vspace{7.7cm} \epsfig{file=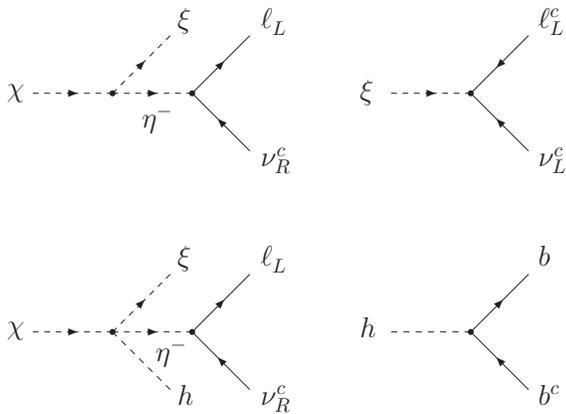, bbllx=3.9cm, bblly=5.9cm,
bburx=13.9cm, bbury=15.9cm, width=7.5cm, height=7.5cm, angle=0,
clip=0} \vspace{-9.5cm} \caption{\label{dm} The dark matter ($\chi$)
decays into the charged leptons ($\ell$), the neutrinos ($\nu$) and
the bottoms ($b$).}
\end{figure}

The recent cosmic-ray experiments
\cite{chang2008,torii2008,adriani2008,aharonian2008,abdo2009}
suggest \cite{mpsv2009} that (i) the TeV-scale dark matter should
mostly annihilate or decay into the leptons with a large cross
section or a long life time; (ii) the dark matter annihilation or
decay should not result in overabundant gamma and neutrino fluxes.
For demonstration, we take the rotation,
\begin{eqnarray}
&&Z_1^{}\left[\langle\phi\rangle\phi+\langle\eta_1^{}\rangle\eta_1^{}+\langle\eta_2^{}\rangle\eta_2^{}\right]
\rightarrow\phi =\left(
\begin{array}{c}
\frac{1}{\sqrt{2}}h+\langle\phi\rangle\\
0\end{array} \right)\,,\nonumber\\
[1mm]
&&Z_1^{}Z_2^{}\left[\langle\phi\rangle\left(\langle\eta_1^{}\rangle\eta_1^{}+\langle\eta_2^{}\rangle\eta_2^{}\right)
-\left(\langle\eta_1^{}\rangle^2_{}+\langle\eta_2^{}\rangle^2_{}\right)\phi\right]\rightarrow\eta_1^{}
\,,\nonumber\\
[2mm] &&Z_2^{}\left[
\langle\eta_1^{}\rangle\eta_2^{}-\langle\eta_2^{}\rangle\eta_1^{}\right]
\rightarrow\eta_2^{}\,,
\end{eqnarray}
where $Z_1^{} =
[\langle\phi\rangle^2_{}+\langle\eta_1^{}\rangle^2_{}
+\langle\eta_2^{}\rangle^2_{}]^{-\frac{1}{2}}_{}$ and $Z_2^{} =
[\langle\eta_1^{}\rangle^2_{}
+\langle\eta_2^{}\rangle^2_{}]^{-\frac{1}{2}}_{}$. Clearly, the
Yukawa couplings of the heavy doublet scalars to the quarks are
highly suppressed by
$\displaystyle{\mathcal{O}(\langle\eta\rangle/\langle\phi\rangle)}$.
Therefore, the dark matter $\chi$ should mostly decay into the
charged singlet scalar $\xi$, the left-handed charged leptons
$\ell_L^{}$, the right-handed neutrinos $\nu_R^{}$ and the physical
Higgs boson $h$. We show the dominant decay channels in Fig.
\ref{dm}. Here we have taken into account that (1) $\xi$ can
uniquely and rapidly decay into the left-handed leptons $\ell_L^{}$
and $\nu_L^{}$ through the Yukawa coupling
$f\xi\overline{\psi_{L}^c}i\tau_2^{}\psi_{L}^{}$; (2) $h$ can
dominantly decay into the bottom quarks. The decay width is given by
\begin{eqnarray}
\Gamma_{\chi}^{}&=&\frac{1}{192\cdot
(2\pi)^3}\Sigma_{i,j=1}^{2}\kappa_i^{\ast}\kappa_j^{}\textrm{Tr}\left(y_i^{}y^\dagger_j\right)
\frac{\langle\phi\rangle^2_{}m_\chi^3}{M_{\eta_i^{}}^2M_{\eta_j^{}}^2}\nonumber\\
[1mm]
&&\times\left[1+\frac{1}{96\cdot\left(2\pi\right)^2_{}}\frac{m_\chi^2}{\langle\phi\rangle^2_{}}\right]\,.
\end{eqnarray}
For giving a numerical example, we choose the parameters considered
before and then determine the life time,
\begin{eqnarray}
\tau_\chi^{} = \frac{1}{\Gamma_\chi^{}}&\simeq & 0.74\times
10^{26}\,\textrm{sec}\,.
\end{eqnarray}
We notice that the SM Higgs boson and then the bottom quark appear
in the final states of the dark matter decay. However, the branching
ratio is small (less than $4\%$) for the present choice of the
parameters. This means the dark matter mostly decays into the
leptons. Compared with the good fitting in \cite{mpsv2009}, we find
that the dark matter decay in our model can induce the desired
positron/electron excesses but avoid the overabundant gamma and
neutrino fluxes.

\vspace{2mm}

The quartic coupling of the dark matter scalar $\chi$ to the SM
Higgs doublet $\phi$, i.e.
$\beta\left(\chi^\dagger_{}\chi\right)\left(\phi^\dagger_{}\phi\right)$,
will result in an elastic scattering of dark matter on nuclei and
hence a nuclear recoil \cite{aht2008}. The spin-independent cross
section of the dark-matter-nucleon elastic scattering would be,
\begin{eqnarray}
\label{dama3} &&\sigma\left(\chi N \rightarrow \chi N\right)=
\frac{\beta^2_{}}{4\pi}\frac{\mu_r^2}{m_h^4
m_\chi^2}f^2_{}m_N^2\nonumber\\
[2mm] && \hskip -.5cm\simeq 10^{-44}_{}\,\textrm{cm}^2_{}\,\times
\frac{\beta^2_{}}{4\pi}\left(\frac{f}{0.3}\right)^2_{}\left(\frac{120\,\textrm{GeV}}{m_h^{}}\right)^4_{}
\left(\frac{4\,\textrm{TeV}}{m_\chi^{}}\right)^2_{}\,,\phantom{xxx}
\end{eqnarray}
which is below the current experimental limit
\cite{angle2007,ahmed2008}, however, reachable for the future
experiments. Here $\mu_{r}^{}=m_\chi^{} m_N^{}/(m_\chi^{} + m_N^{})$
is the nucleon-dark-matter reduced mass, $m_h^{}$ is the mass of the
physical Higgs boson $h$ (The mixing between $h$ and $h'$ (from
$\sigma$) is negligible for
$\langle\sigma\rangle\gg\langle\phi\rangle$.), the factor $f$ with a
central value $f=0.30$ \cite{aht2008} parameterizes the Higgs to
nucleons coupling from the trace anomaly, $fm_{N}^{}\equiv\langle
N|\sum_q^{}m_q^{}\bar{q}q|N\rangle$.

\vspace{2mm}

In this paper we connect the neutrino masses with the origin of the
dark matter relic density and the baryon asymmetry in the present
Universe. In our model, the dark matter relic density is a dark
matter asymmetry because the thermal relic density of dark matter is
negligible in the presence of the fast annihilation between the dark
matter and dark antimatter. This dark matter asymmetry originate
simultaneously with a lepton asymmetry, which can explain the baryon
asymmetry via the sphaleron process, and given by the amount of CP
violation in out-of-equilibrium decays of heavy scalars. The dark
matter mostly decays into the leptons to generate the
positron/electron excesses without the overabundant gamma and
neutrino fluxes so that we can explain the results from the recent
cosmic-ray experiments.

\vspace{5mm}

\noindent \textbf{Acknowledgement}: We thank Xiao-Jun Bi and Goran
Senjanovi$\rm\acute{c}$ for helpful discussions. US thanks the
Department of Physics and the McDonnell Center for the Space
Sciences at Washington University in St. Louis for inviting him as
Clark Way Harrison visiting professor. XZ is supported in part by
the National Natural Sciences Foundation of China under grants
10533010 and 10675136, and by the Chinese Academy of Sciences under
grant KJCX3-SYW-N2.

\end{document}